\documentclass[a4paper,fleqn,10pt]{scrartcl}
\pdfoutput=1

\begingroup\expandafter\expandafter\expandafter\endgroup
\expandafter\ifx\csname pdfsuppresswarningpagegroup\endcsname\relax
\else
\fi

\usepackage{microtype}
\usepackage{authblk}

\usepackage{amsmath}
\usepackage{amssymb}
\usepackage{commath}
\usepackage{array}
\usepackage{calc}
\usepackage{siunitx}
\usepackage{booktabs}
\usepackage{enumitem}
\usepackage{multirow}
\usepackage{graphicx}
\graphicspath{{figures/}}
\usepackage{xspace}
\usepackage{
  pgf,
  tikz}
\usetikzlibrary{
  patterns,
  shapes.multipart,
  arrows,
  trees,
  scopes,
  decorations.pathreplacing,
  decorations.pathmorphing,
  decorations.markings,
  decorations.text,
  positioning,
  calc
}
\usepackage{mciteplus}
\usepackage[pdfborder={0 0 0}]{hyperref}
\usepackage[format=hang,labelfont=bf,hypcap=true]{caption}
\usepackage{subfig}
\usepackage{relsize}

\let\spreprint\empty
\newcommand{\preprint}[1]{\def\spreprint{\protect#1}}
\let\sinstitute\empty

\makeatletter
\renewcommand{\maketitle}{\begingroup
  \null\thispagestyle{empty}%
    \ifx\spreprint\empty
      \vskip 5ex
    \else
      \flushright\large\spreprint\vskip 2ex
    \fi
    \vskip 5ex
    \flushleft
      {\sffamily\bfseries\huge\@title}\vskip 2ex
      \@author\vskip 2ex
      \ifx\sinstitute\empty
      \else
        {\small\sinstitute}
      \fi
    \vskip 5ex
  \endgroup
}
\makeatother
\renewenvironment{abstract}{\begin{center}
  {\large\sffamily\bfseries Abstract: }
  \begin{minipage}[t]{0.75\textwidth}
}{\end{minipage}\end{center}\vskip 10ex}

\newcommand{\mytextwidthfigure}[3]{
  \begin{figure}[#1]
    \begin{center}
      #2\\
      \parbox[t]{\textwidth}{\caption{#3}}
    \end{center}
  \end{figure}
}

\numberwithin{equation}{section}

\newcommand{\shortequal}{\ensuremath{\!\!\!=\!\!\!}}

\newcommand{\bea}{\begin{align}}
\newcommand{\eea}{\end{align}}
\newcommand{\beq}{\begin{equation}}
\newcommand{\eeq}{\end{equation}}

\newcommand{\bs}{\begin{split}}
\newcommand{\es}{\end{split}}

\newcommand{\bi}{\begin{itemize}}
\newcommand{\ei}{\end{itemize}}
\newcommand{\bc}{\begin{center}}
\newcommand{\ec}{\end{center}}
\newcommand{\bac}{\begin{array}{c}}
\newcommand{\bacc}{\begin{array}{cc}}
\newcommand{\ea}{\end{array}}

\def\spa#1.#2{\langle#1\,#2\rangle}
\def\spb#1.#2{[#1\,#2]}

\newcommand{\pt}{\ensuremath{p_T}\xspace}

\newcommand{\pT}{\pt}

\newcommand{\alphaS}{\ensuremath{\alpha_\text{s}}\xspace}

\newcommand{\muR}{\ensuremath{\mu_\mathrm{R}}}
\newcommand{\muF}{\ensuremath{\mu_\mathrm{F}}}
\newcommand{\muQ}{\ensuremath{\mu_\mathrm{Q}}}

\newcommand{\NLO}{\ensuremath{\mathrm{NLO}}\xspace}
\newcommand{\NNLO}{\ensuremath{\mathrm{NNLO}}\xspace}

\newcommand{\sla}[1]{\ensuremath{{#1\kern-0.45em/}}}

\newcommand{\SB}{\ensuremath{S/B}}

\newcommand{\antikT}{anti-$k_T$}

\newcommand{\SecRef}[1]{Section~\ref{#1}\xspace}

\newcommand{\FigureRef}[1]{Figure~\ref{#1}\xspace}
\newcommand{\FigRef}[1]{Fig.~\ref{#1}\xspace}

\newcommand{\TabRef}[1]{Tab.~\ref{#1}\xspace}

\newcommand\LHC{L\protect\scalebox{0.8}{HC}\xspace}

\newcommand{\MCatNLO}{M\protect\scalebox{0.8}{C}@N\protect\scalebox{0.8}{LO}\xspace}
\newcommand{\SMCatNLO}{S-M\protect\scalebox{0.8}{C}@N\protect\scalebox{0.8}{LO}\xspace}
\newcommand{\POWHEG}{P\protect\scalebox{0.8}{OWHEG}\xspace}

\newcommand{\MEPSatNLO}{M\scalebox{0.8}{E}P\scalebox{0.8}{S}@N\protect\scalebox{0.8}{LO}\xspace}

\newcommand{\CutTools}{C\protect\scalebox{0.8}{UT}T\protect\scalebox{0.8}{OOLS}\xspace}

\newcommand{\Collier}{C\protect\scalebox{0.8}{OLLIER}\xspace}

\newcommand{\OpenLoops}{O\protect\scalebox{0.8}{PEN}L\protect\scalebox{0.8}{OOPS}\xspace}

\newcommand{\OneLoop}{O\protect\scalebox{0.8}{NE}L\protect\scalebox{0.8}{OOP}\xspace}
\newcommand{\MCFM}{M\protect\scalebox{0.8}{C}F\protect\scalebox{0.8}{M}\xspace}

\newcommand{\Sherpa}{S\protect\scalebox{0.8}{HERPA}\xspace}
\newcommand{\Comix}{C\protect\scalebox{0.8}{OMIX}\xspace}

\newcommand{\Amegic}{A\protect\scalebox{0.8}{MEGIC}\xspace}
\newcommand{\CSS}{C\protect\scalebox{0.8}{SS}\xspace}

\newcommand{\LHAPDF}{L\protect\scalebox{0.8}{HAPDF}}

\newcommand{\APFEL}{A\protect\scalebox{0.8}{PFEL}\xspace}
\newcommand{\NNPDF}{N\scalebox{0.8}{NPDF}}
\hypersetup{pdfauthor={Enrico Bothmann, Frank Krauss, Marek Sch{\"o}nherr},
  pdftitle={Single top-quark production with Sherpa}}

\preprint{CERN-TH-2017-234\\EDINBURGH-17-22\\IPPP/17/83\\MCNET-17-17}

\author[1]{Enrico Bothmann}
\affil[1]{Higgs Centre for Theoretical Physics, University of Edinburgh, Edinburgh EH9 3FD, UK}

\author[2]{Frank Krauss}
\affil[2]{Institute for Particle Physics Phenomenology, Durham University, Durham DH1 3LE, UK}

\author[3]{Marek Sch{\"o}nherr}
\affil[3]{TH Division, Physics Department, CERN, CH--1211 Geneva 23, Switzerland}

\title{Single top-quark production with \Sherpa}

\begin{document}
\maketitle
\begin{abstract}
  We present results at next-to-leading order accuracy in QCD for single
top-quark production in the $t$, $s$ and $tW$ channels at the \LHC at a
centre-of-mass energy of \SI{8}{\TeV}, obtained
with the \Sherpa event generator.  We find them in very good agreement 
with measured values and quantify their theory uncertainties. Uncertainties 
stemming from the choice between the four- and the five-flavour scheme are
found to be typically of the order of 5--\SI{10}{\percent} over large ranges of
phase space.
We discuss the impact of parton distribution functions, and in particular
of the bottom PDF.  We also show how different cuts on QCD radiation
patterns improve the signal to background ratio in realistic fiducial 
volumes.

\end{abstract}

\section{Introduction}

The production of single-top quarks is an important source of backgrounds in
searches for new physics~\cite{Tait:2000sh,AguilarSaavedra:2008gt,Gao:2011fx},
but it is also a signal in its own right~\cite{Aad:2012ux,Chatrchyan:2014tua,
  Khachatryan:2014iya,Aad:2015eto,Aad:2015upn,Khachatryan:2016ewo,Aaboud:2017pdi}
since it allows the direct determination of the Cabibbo-Kobayashi-Maskawa
matrix element $|V_{tb}|$~\cite{%
Alwall:2006bx,Aaltonen:2010jr,Khachatryan:2014iya}.
By far and large it has become customary to distinguish three modes for
single-top production, differentiated by the role played by the $W$ boson, namely
$s$-channel production ($q\bar{q}'\to t\bar{b},\,\bar{t}b$ at Born level),
$t$-channel production ($q\bar{b}\to q'\bar{t}$ and $\bar{q}b\to \bar{q}'t$ at
Born level), and $tW$-associated production, ($gb\to tW^-$ and
$g\bar{b}\to \bar{t}W^+$ at Born-level). Fixed-order
predictions at next-to-leading accuracy in QCD (NLO) have been presented for the $s$-channel 
in~\cite{Smith:1996ij,Harris:2002md}, for the $t$-channel
in~\cite{Bordes:1994ki,Harris:2002md}, and for $tW$-associated production
in~\cite{Giele:1995kr}. Monte-Carlo simulations that
are accurate to NLO have been constructed for all
three channels with the \MCatNLO method~\cite{Frixione:2005vw,Frixione:2008yi}
and with the \POWHEG method~\cite{Alioli:2009je,Re:2010bp}.
Four-flavour scheme variants of these results have been presented and compared
with the five-flavour scheme ones in~\cite{Campbell:2009ss,Frederix:2012dh}.
Results for the production and subsequent decay of
single top-quarks at NLO precision for the $s$ and the $t$ channel have been
presented in~\cite{Campbell:2004ch}, as part of the \MCFM package.
Furthermore, the cross section for the dominant $t$-channel mode has been
calculated up to \NNLO in QCD~\cite{Brucherseifer:2014ama}.
Electroweak corrections to single-top production have been discussed
in~\cite{Bardin:2010mz}.

In this publication, we present results obtained with the \Sherpa event
generation framework~\cite{Gleisberg:2008ta} for single-top production in
all three channels.  After a short description of the generation setups below,
\SecRef{Sec::Setups}, we will contrast our results with experimentally 
measured data in \SecRef{Sec::Validation}. In this section we also 
further investigate theory uncertainties on 
typical distributions for the two production channels, with some emphasis
on the use of the four- or five-flavour scheme
for $t$-channel production. We also comment 
on the impact of different bottom parton distribution functions (PDFs) on selected 
observables.  Finally, in \SecRef{Sec::Results}, we will investigate
signal-\-to-\-background ratios, especially for $t$-channel production, 
and how they can be improved through cuts on light jets, before 
summarising our findings in \SecRef{sec:summary}.

\section{Setup}
\label{Sec::Setups}

We calculate all three single-top production channels, 
the $t$-, the $s$- and the $tW$-channel at a 
centre-of-mass energy of $\sqrt{s}=\SI{8}{\TeV}$, using the 
\MCatNLO technique \cite{Frixione:2002ik,Frixione:2005vw} in 
the variant implemented in \Sherpa, \SMCatNLO 
\cite{Hoeche:2011fd,Hoeche:2012ft,Hoche:2012wh}.  
Up to NLO QCD the $t$- and $s$-channel process are unequivocally 
defined by the presence of a $t$- and $s$-channel $W$~boson, 
respectively. However, the $tW$-channel overlaps at NLO
with $t\bar{t}$ production. 
This overlap is resolved using the diagram removal 
technique of \cite{Frixione:2008yi}, excluding doubly 
resonant diagrams.\footnote{A fully consistent treatment of the $tW$-channel at
\NLO would involve the calculation of $W^+W^-b\bar{b}$ production, with the
$tW$-channel simply being the singly-resonant contribution for observables 
that are inclusive in one of the $b$-quarks~\cite{Denner:2010jp,
  Denner:2012yc,Frederix:2013gra,Cascioli:2013wga,Jezo:2016ujg}.}
To assess uncertainties due to the flavour-scheme, we further calculate
both the $t$- and $s$-channel processes in the five- and four-flavour
schemes.  The $tW$-channel is calculated in the five-flavour scheme only, 
as the afore mentioned ambiguous and gauge-dependent 
removal of resonant $t\bar{t}$ production is already present at LO in 
the four-flavour scheme~\cite{Frederix:2013gra,Cascioli:2013wga}. 
The dominant background processes for the analysis in \SecRef{Sec::Results}, 
$t\bar{t}$ production and $W$-boson production in association 
with at least one light- and one $b$-jet,
also use the \MCatNLO technique.

Tree-level matrix elements and subtraction terms in the 
Catani-Seymour dipole formalism~\cite{Catani:1996vz,Catani:1996jh,
  Gleisberg:2007md} are generated using the
\Amegic~\cite{Krauss:2001iv} and \Comix~\cite{Gleisberg:2008fv} matrix-element 
generators.  One-loop matrix elements
are taken from the \OpenLoops library~\cite{Cascioli:2011va}, relying on 
\Collier~\cite{Denner:2014gla}, \CutTools~\cite{Ossola:2007ax} and 
\OneLoop~\cite{vanHameren:2010cp}.  
All partons are evolved from their high scales at production to low scales 
through a Catani-Seymour dipole shower, \CSS~\cite{Schumann:2007mg}.

Top quarks are produced on-shell with $m_t = \SI{172.5}{\GeV}$ in the
zero-width approximation, before they are decayed into a $W$ boson and a
bottom quark.  As in~\cite{Hoche:2014kca} the kinematics of the decay are
adjusted a posteriori to the physical width 
of the top quark and the $W$ boson by redistributing their masses according 
to the respective Breit-Wigner distribution. 
These $W$ bosons are further decayed, either leptonically ($\ell=e,\mu,\tau$), 
or semi-leptonically in the case of the $t\bar{t}$ background simulation.  
$W$ bosons that are not part of the top-quark decay chain, e.g.\ 
in the $tW$ channel, are decayed hadronically.  
The full decay chain accounts for spin-correlations and intermediate 
QCD and QED corrections through either constrained parton-shower 
evolution or soft-photon resummation 
in the YFS-scheme \cite{Yennie:1961ad,Schonherr:2008av}. 
The branching ratios are correctly taken into account throughout, with 
the exception of the total cross sections in \SecRef{Sec::Validation} 
which are quoted for inclusive single-top production.

We further include the simulation of multiple parton interactions 
according to the method laid down in~\cite{Sjostrand:1987su}. Its 
\Sherpa implementation has been described in~\cite{Alekhin:2005dx}. 
A hadronisation simulation \cite{Winter:2003tt} and hadron decays, 
including both hadronic and leptonic $\tau$ decays and supplemented 
by higher order QED corrections~\cite{Schonherr:2008av}, are also 
included, in order to arrive at a particle-level calculation.

\begin{table}[t]
\centering
\begin{tabular}{lcc}
\toprule
process                            & scales ($N_f=5$) & scales ($N_f=4$) \\
\midrule
\addlinespace[0.5em]
$pp\to tq/\bar{t}q$ (t-channel)  	   & $-Q_W^2$ & $\muF^2=\muQ^2=m_\text{T}^{t\,2}$, $\muR^2=m_\text{T}^{b\,2}$ \\
\addlinespace[0.5em]
$pp\to t\bar{b}/\bar{t}b$ (s-channel)  	   & $Q_W^2$        & $Q_W^2$ \\
\addlinespace[0.5em]
$pp\to tW^-/\bar{t}W^+$ (associated production) & $m_\text{T}^{t\,2}$ & ---\\
\bottomrule
\end{tabular}\\[2mm]
\caption{Scale choices for all three single-top production channels.}
\label{tab:scales-top}
\end{table}

\begin{table}[t]
\centering
\begin{tabular}{lc}
\toprule
process                            & scales \\
\midrule
\addlinespace[0.5em]
$pp\to Wjb$ 		& $\frac{1}{4}\,H_\mathrm{T}^{\prime\,2}$ \\
\addlinespace[0.5em]
$pp\to t\bar{t}$	& $-\frac{1}{1/\hat{s}+1/\hat{t}+1/\hat{u}}$ \\
\addlinespace[0.5em]
\bottomrule
\end{tabular}\\[2mm]
\caption{Scale choices for $W$+jets and top-pair production.}
\label{tab:scales-bkgrd}
\end{table}

The hard interaction and its matching to the parton shower are 
characterised by three scales: the renormalisation scale $\muR$, 
the factorisation scale $\muF$, and the resummation scale $\muQ$.
The latter can be identified as the parton-shower starting scale. 
For the different single-top signal channels, the scales are set 
as listed in \TabRef{tab:scales-top}.
For the $t\bar{t}$- and $W$-boson backgrounds, the scale choices are
listed in \TabRef{tab:scales-bkgrd}. For $t\bar{t}$, the
clustering algorithm of 
the \MEPSatNLO multijet merging method 
\cite{Hoeche:2012yf,Gehrmann:2012yg,Hoeche:2013mua,Hoeche:2014qda} 
determines the emission scales up to the scale of the $2\to2$ core
process. Hence, we list this core scale for $t\bar{t}$ in
\TabRef{tab:scales-top}.
The PDFs for our central value are given by the \NNPDF\,3.0 set at 
\NLO~\cite{Ball:2014uwa} in the appropriate flavour number scheme, 
interfaced through \LHAPDF\,6~\cite{Buckley:2014ana}.  
The electroweak couplings $\alpha$ are evaluated with the $G_\mu$ scheme
as suggested in~\cite{Andersen:2014efa}. All other input parameters are
detailed in \TabRef{tab:inputs}.
%
%
\begin{table}[tb]
  \begin{center}
    \begin{tabular}{rclrcl}
      $G_\mu$ & \shortequal & $1.16639\cdot 10^{-5}~\text{GeV}^2$ & \qquad\qquad & &\\
      $m_W$ & \shortequal & $80.385~\text{GeV}$  & $\Gamma_W$ & \shortequal & $2.085~\text{GeV}$ \\
      $m_Z$ & \shortequal & $91.1876~\text{GeV}$ & $\Gamma_Z$ & \shortequal & $2.4952~\text{GeV}$ \\
      $m_b$ & \shortequal & $4.75~\text{GeV}$    & $\Gamma_b$ & \shortequal & $0$ \\
      $m_t$ & \shortequal & $172.5~\text{GeV}$   & $\Gamma_t$ & \shortequal & $1.47015~\text{GeV}$ 
    \end{tabular}
  \end{center}
  \caption{
    Numerical values of all input parameters. In calculations where 
    a given particle is present as a final state its width is set 
    to zero. The value listed above is then used in the redistribution 
    of its kinematics in the generation of its factorised decay. 
    The bottom-quark mass is only used in four-flavour scheme calculations.
  }
  \label{tab:inputs}
\end{table}

Theory uncertainties are generated on-the-fly using the internal reweighting of
\Sherpa~\cite{Bothmann:2016nao}.  Where scale variations are given, they amount
to the envelope over a 7-point scale variation, independently multiplying 
\muF\ and \muR\ by factors of two and one half, but not allowing variations
where one scale is scaled up and the other one down.
Where clustering is used in a calculation, only the core process scale is
affected by the variation,
the clustering scales are kept at their central values.  To estimate PDF
errors, the variations for the \NNPDF\ replicas are combined as a statistical
sample~\cite{Ball:2014uwa}.  To vary \alphaS, we generate results for PDF
variations that are fitted using different input values for $\alphaS(m_Z)$. The
central value for it is 0.118, and the variations are 0.117 and 0.119. The
\alphaS\ error is then given as the envelope over the three corresponding predictions.
PDF and \alphaS\ variations are not applied to the parton shower.
All three sources of uncertainties, the scale, PDF and $\alphaS$ uncertainties, 
are either added linearly or given individually, if not specified otherwise.

\section{Total and fiducial cross sections and uncertainties}
\label{Sec::Validation}

\newlength{\verticallinewidth}
\verticallinewidth0.25em
\makeatletter
\g@addto@macro{\endtabular}{\rowfont{}}
\makeatother
\newcommand{\rowfonttype}{}
\newcommand{\rowfont}[1]{
   \gdef\rowfonttype{#1}#1%
}
\newcolumntype{L}{>{\rowfonttype}l}
\newcolumntype{C}{>{\rowfonttype}c}
\begin{table}[p]
\centering
\begin{tabular}{@{}CCC@{\qquad}L@{\quad}L@{\quad}L@{\quad}L@{\qquad}L@{\quad}L@{\quad}L@{\quad}L@{\qquad}LL@{\qquad}LL@{}}
\toprule
  & & & \multicolumn{4}{l}{\Sherpa ($N_f=5$)}& \multicolumn{4}{l}{\Sherpa ($N_f=4$)} &
  \multicolumn{2}{l}{\;ATLAS} & \multicolumn{2}{l}{\;CMS} \\
  \rowfont{\footnotesize}%
  & & & &             &         &     & &             &         &     &
\multicolumn{2}{l}{%
  \footnotesize\cite{Aad:2015eto,Aaboud:2017pdi,Aad:2015upn}} &
\multicolumn{2}{l}{%
  \footnotesize\cite{Khachatryan:2014iya,Khachatryan:2016ewo,Chatrchyan:2014tua}} \\
& & & & $\mu_{R,F}$ & \alphaS & PDF & & $\mu_{R,F}$ & \alphaS & PDF &
& tot. & & tot. \\
\midrule
\rowfont{\normalsize}%
\multirow{4}{\verticallinewidth}{\parbox[c][8em][c]{\verticallinewidth}{\rotatebox[origin=c]{90}{$t$-channel}}} &
\multirow{2}{\verticallinewidth}{\parbox[c][3em][c]{\verticallinewidth}{\rotatebox[origin=c]{90}{tot.}}} &
$t$ &
58.3 & $^{+1.8}_{-1.4}$ & $^{+0.4}_{-0.6}$ & $\pm{0.7}$ &  
58.3 & $^{+2.8}_{-3.6}$ & $^{+0.6}_{-0.7}$ & $\pm{0.6}$ &  
$56.7$ & $^{+4.3}_{-3.8}$ & 53.8 & $\pm4.7$                
\\ \addlinespace[1.0em]
&
&
$\bar{t}$ &
32.1 & $^{+1.0}_{-0.8}$ & $^{+0.3}_{-0.4}$ & $\pm{0.5}$ &
34.7 & $^{+3.5}_{-3.0}$ & $^{+0.5}_{-0.5}$ & $\pm{0.5}$ &
$32.9$ & $^{+3.0}_{-2.7}$ & 27.6 & $\pm4.0$
\\ \addlinespace[1.5em]
&
\multirow{2}{\verticallinewidth}{\parbox[c][3em][c]{\verticallinewidth}{\rotatebox[origin=c]{90}{fid.}}} &
$t$ &
9.30 & $^{+0.36}_{-0.29}$ & $^{+0.06}_{-0.10}$ & $\pm{0.11}$ &
9.35 & $^{+0.63}_{-0.69}$ & $^{+0.09}_{-0.11}$ & $\pm{0.10}$ &
$9.78$ & $\pm0.57$ & --- &
\\ \addlinespace[1.0em]
&
&
$\bar{t}$ &
5.09 & $^{+0.21}_{-0.17}$ & $^{+0.04}_{-0.06}$ & $\pm{0.08}$ &
5.72 & $^{+0.71}_{-0.57}$ & $^{+0.08}_{-0.09}$ & $\pm{0.08}$ &
$5.77$ & $\pm0.45$ & --- &
\\ \addlinespace[1.5em]
\multirow{2}{\verticallinewidth}{\parbox[c][3em][c]{\verticallinewidth}{\rotatebox[origin=c]{90}{$s$-ch.}}} &
\multirow{2}{\verticallinewidth}{\parbox[c][3em][c]{\verticallinewidth}{\rotatebox[origin=c]{90}{tot.}}} &
$t$ &
3.31 & $^{+0.09}_{-0.07}$ & $^{+0.01}_{-0.02}$ & $\pm{0.06}$ &         
3.26 & $^{+0.09}_{-0.07}$ & $^{+0.01}_{-0.02}$ & $\pm{0.06}$ &         
\multirow{2}{*}{$4.8$} & \multirow{2}{*}{$^{+1.8}_{-1.6}$} &           
\multirow{2}{*}{$13.4$} & \multirow{2}{*}{$\pm7.3$}                    
\\ \addlinespace[1.0em]
&
&
$\bar{t}$ &
1.89 & $^{+0.05}_{-0.04}$ & $^{+0.01}_{-0.01}$ & $\pm{0.04}$ &         
1.87 & $^{+0.05}_{-0.04}$ & $^{+0.01}_{-0.01}$ & $\pm{0.04}$ &         
  & &
\\ \addlinespace[0.75em]
\multirow{2}{\verticallinewidth}{\parbox[c][3em][c]{\verticallinewidth}{\rotatebox[origin=c]{90}{$tW$-ch.}}} &
\multirow{2}{\verticallinewidth}{\parbox[c][3em][c]{\verticallinewidth}{\rotatebox[origin=c]{90}{tot.}}} &
$t$ &
12.3 & $^{+0.8}_{-0.7}$ & $^{+0.2}_{-0.2}$ & $\pm{0.4}$ &         
\multicolumn{4}{l}{---} &                                         
\multirow{2}{*}{$23.0$} & \multirow{2}{*}{$^{+3.7}_{-3.9}$} &     
\multirow{2}{*}{$23.4$} & \multirow{2}{*}{$\pm5.4$}               
\\ \addlinespace[1.0em]
&
&
$\bar{t}$ &
12.3 & $^{+0.8}_{-0.7}$ & $^{+0.2}_{-0.2}$ & $\pm{0.4}$ &         
\multicolumn{4}{l}{---} &                                         
  & &
\\ \addlinespace[0.25em] \bottomrule
%
\end{tabular}
\caption{\label{tab:xs}%
Total and fiducial single-top production cross 
sections in picobarn.  The omitted statistical errors for 
the \Sherpa results are at least an order of magnitude 
smaller than their scale uncertainties. All \Sherpa 
results are generated at \MCatNLO accuracy. The fiducial 
cross sections are defined by the cuts given in 
\cite{Aaboud:2017pdi}, cf.\ \SecRef{Sec::Results}. The 
quoted experimentally measured values only give the total 
uncertainty. For the $s$- and the $tW$-channel, experimental 
results are only available for the combination of top and 
anti-top production.}
\end{table}

\mytextwidthfigure{p}{%
  \includegraphics{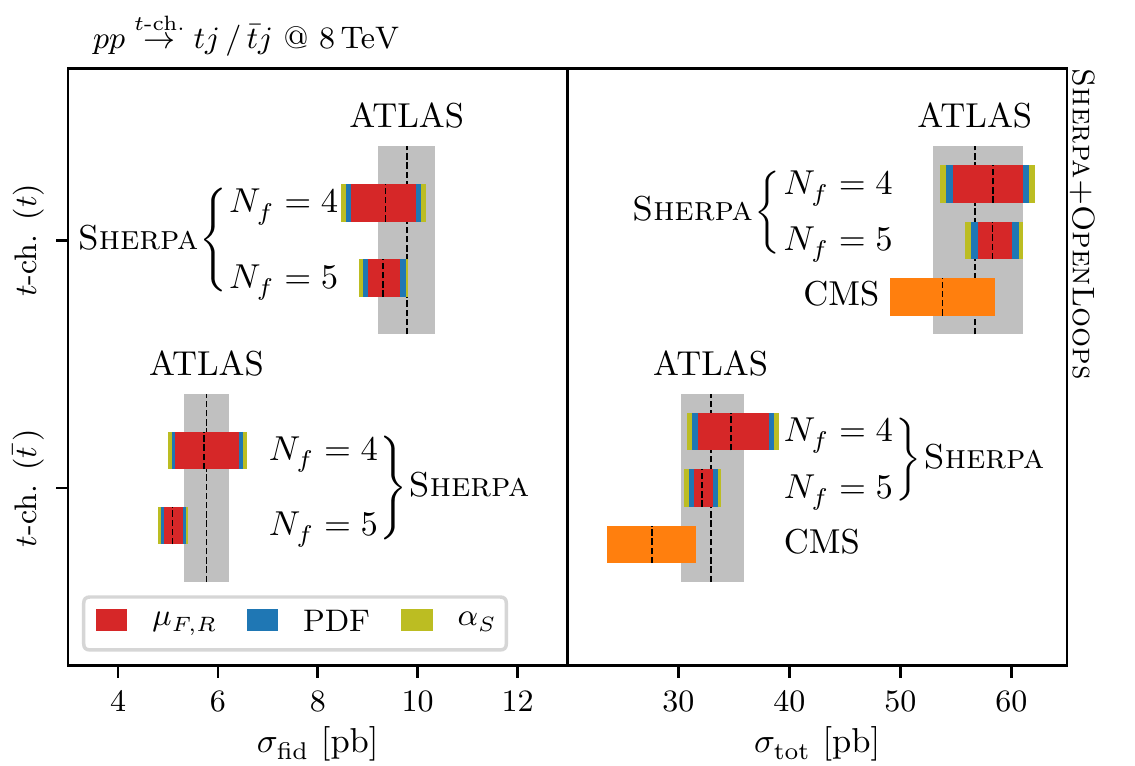}%
  \hfill%
  \includegraphics{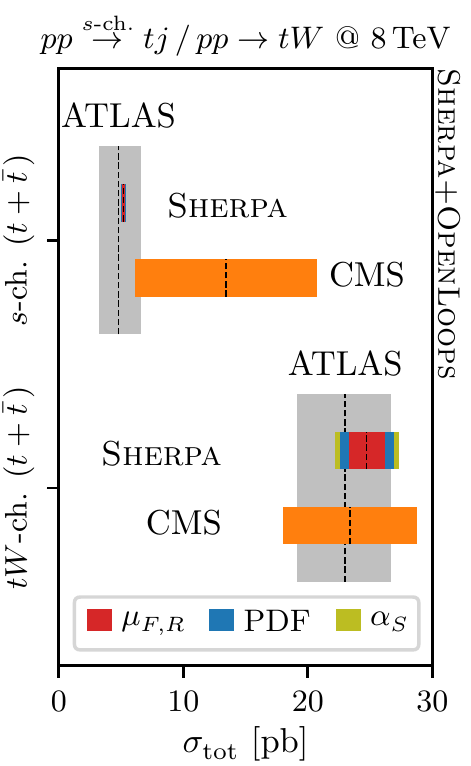}%
  }{\label{fig:xs}Depiction of the cross sections from \TabRef{tab:xs}.  The uncertainties
  for the \Sherpa results are displayed staggered, i.e.\ the total
  width of the \Sherpa band corresponds to the scale uncertainty (red), 
  the PDF uncertainty (blue) and the $\alphaS$ uncertainty (yellow) added linearly.
  Only the $N_f=5$ \Sherpa result is shown for the $s$-channel, because the
  $N_f=4$ result is nearly identical.  All \Sherpa results are calculated at
  \MCatNLO.}

In this section we compare our results with recent measurements at the
\SI{8}{\TeV} LHC~\cite{Aad:2015eto,Aaboud:2017pdi,Aad:2015upn,Khachatryan:2014iya,
  Khachatryan:2016ewo,Chatrchyan:2014tua}.  Inclusive total and fiducial
cross sections for both $t$- and $s$-channel top- and antitop-production
as well as for associated $tW^+$ and $\bar{t}W^-$ production are
listed in~\TabRef{tab:xs} and visualised in \FigRef{fig:xs}.  We find good
agreement between the data and our predictions in all three
channels for both the five- and the four-flavour schemes.
Additionally, we compare our $t$-channel computation for the 
reconstructed top-quark transverse momentum and the leading 
light-jet rapidity with ATLAS data \cite{Aaboud:2017pdi} in
\FigRef{fig:distributions}.
Again, we can establish good agreement between our simulation and data.

\mytextwidthfigure{btp}{%
  \includegraphics[width=0.47\textwidth]{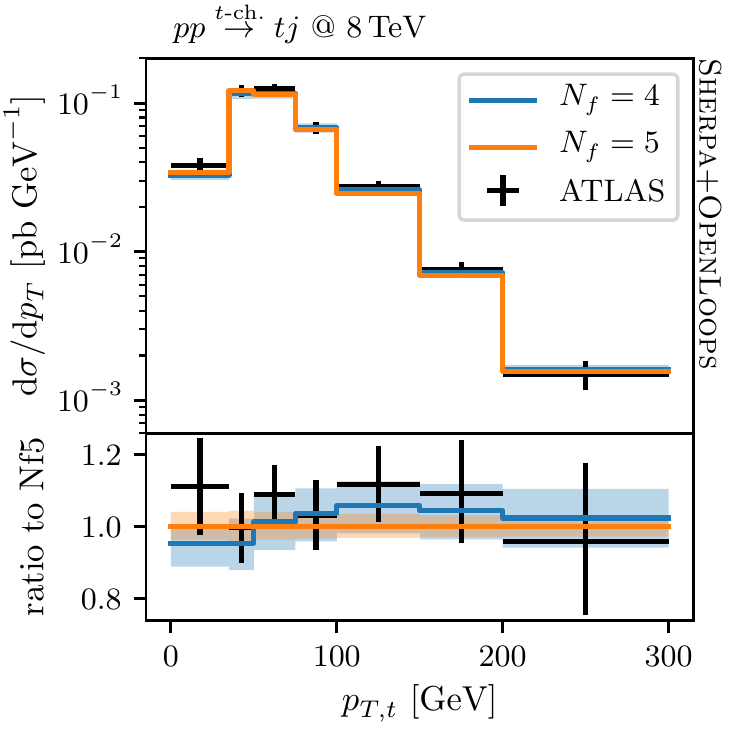}%
  \hfill
  \includegraphics[width=0.47\textwidth]{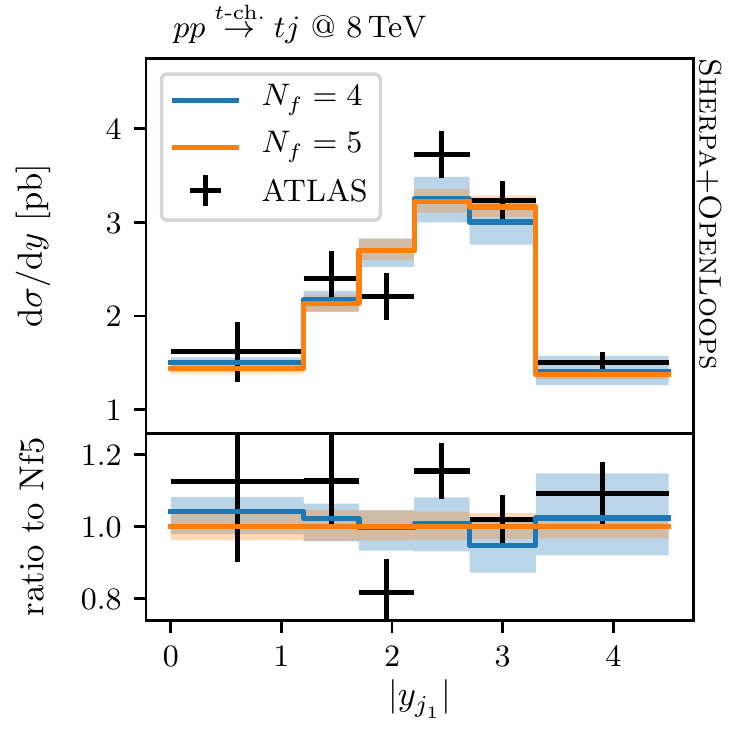}
  }{\label{fig:distributions}Comparison of \Sherpa \MCatNLO predictions with
  ATLAS data~\cite{Aaboud:2017pdi} for the top-quark transverse momentum
  $p_{T,t}$ and the light-jet rapidity $y_{j_1}$ in $t$-channel single-top
  production.  The \Sherpa uncertainty
  consists of the statistical, the \alphaS, the PDF and the (dominating) scale
  uncertainty, all added in quadrature.}

\mytextwidthfigure{p}{%
  \includegraphics[width=0.93\textwidth]{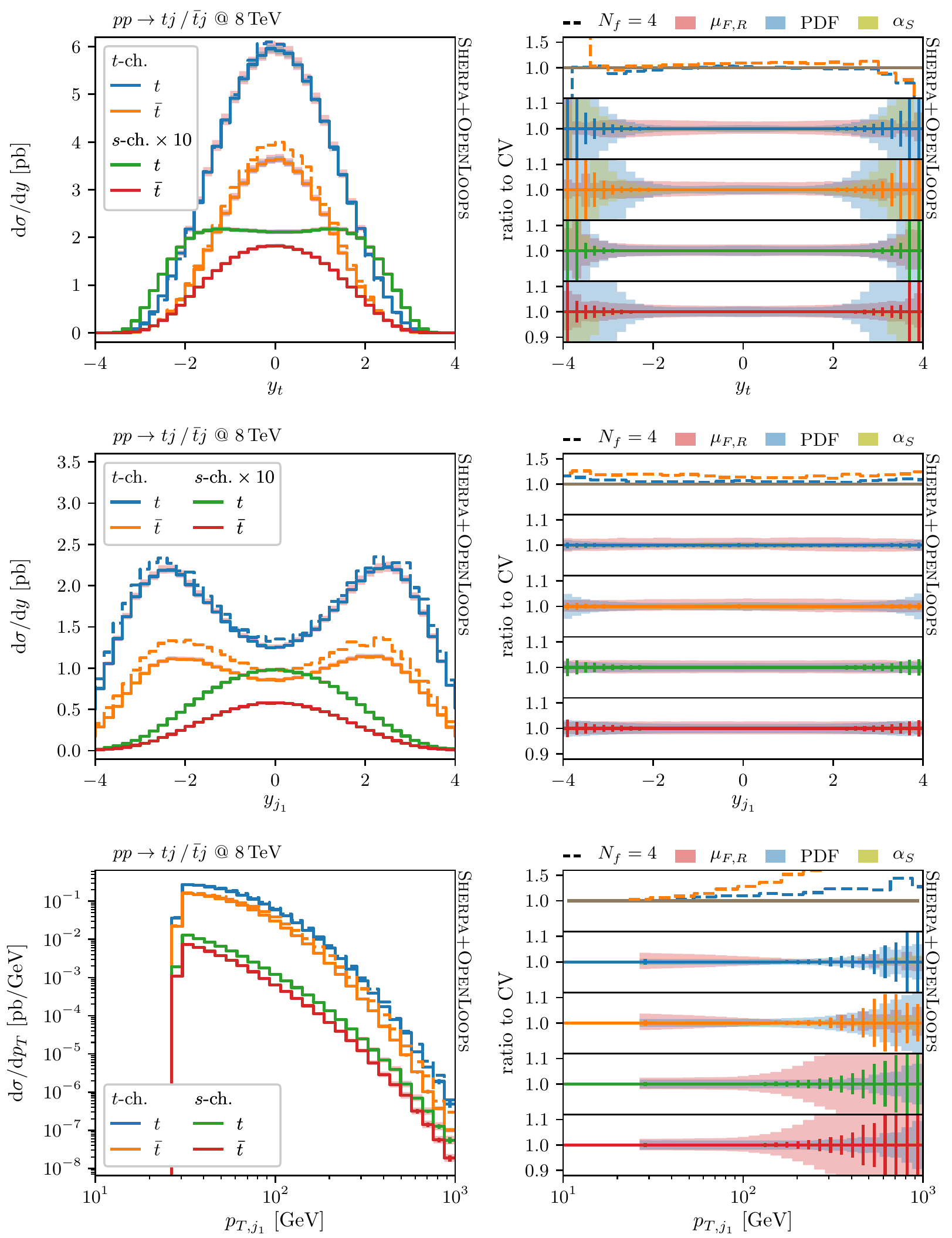}%
  }{\label{fig:s-vs-t}The plots in the left panel show the distribution 
  of the top and antitop quark rapidity $y_t$ (top) as well as the leading jet 
  rapidity $y_{j_1}$ (centre) and its transverse momentum $p_{T,j_1}$ (bottom)
  for leptonic $t$- and 
  $s$-channel single-top production in the five-flavour scheme. The plots
  in the right panel detail their respective uncertainties stemming from 
  the choice for the scales (red band), parton distributions (blue band) and
  the value of $\alphaS$ (yellow band). Each such uncertainty budget is 
  shown separately for each channel: $t$-channel top (blue) and anti-top 
  (orange), and $s$-channel top (green) and anti-top (red) production. 
  The additional panel at the top of each uncertainty breakdown shows 
  the ratio of the four-flavour scheme calculation (dashed) to the 
  corresponding five-flavour calculation (solid) for $t$-channel 
  production. Note that the $s$-channel rapidity distributions have been 
  scaled by a factor of ten.}

In a next step, to further investigate the behaviour of our calculations 
and their associated uncertainties, we compare in \FigRef{fig:s-vs-t} inclusive
$t$- and $s$-channel production with leptonic decays in the five-flavour 
scheme. No acceptance cuts are applied. 
We separately detail the uncertainties stemming from scale variations, 
the parton distributions, and the value of the strong coupling for top 
and anti-top production in both channels. 
For $t$-channel production, they are contrasted with the difference 
of the five- and four-flavour schemes, which we find to be in good agreement.
Of course, both schemes lead to almost identical results in the $s$-channel
production as neither $\alphaS$ nor the bottom PDF appear.
The top and anti-top quark in the $t$~channel are produced centrally, 
and the uncertainties are dominated by the renormalisation and 
factorisation scale variations. PDF uncertainties only become 
relevant beyond rapidities of $|y_t|>2$, which have little 
relevance in the 8\,TeV measurements. Both the five- and four-flavour 
calculations agree on the level of a few percent throughout the entire rapidity 
range, with deviations being slightly larger for anti-top
production. The production via $s$-channel is less central, but otherwise 
exhibits a very similar structure with respect to the uncertainties. 
Slightly larger differences can be observed in the leading light-jet 
rapidity $y_{j_1}$.
It is produced predominantly at large rapidities in 
$t$-channel production, while it is produced centrally 
in the $s$-channel. The uncertainties are entirely dominated 
by scale variations throughout the entire observable 
range. As before, the differences between the five- and four-flavour schemes 
are small, but larger than for $y_t$. Also, they are again larger 
for anti-top production.
The last quantity we assess is the leading-jet transverse momentum $p_{T,j_1}$.
As the $\pT$ increases in $t$-channel production, the uncertainties 
become dominated by the PDF uncertainties. Conversely, in 
$s$-channel production the scale uncertainty rapidly increases as 
the $\pT$ increases, dominating the total uncertainty budget. 
The difference between five-flavour scheme and four-flavour scheme 
predictions is most pronounced in this observable and increases to 
a \SI{10}{\percent} at $\pt=200\,\text{GeV}$ for $t$-channel top productions and 
\SI{40}{\percent} for anti-top production. This difference originates in the 
different interplay of the up(down)-quark and (anti)bottom-quark PDF in the 
dominant five-flavour (anti)top quark production channel, and
the up(down)-quark and the gluon PDF in the dominant four-flavour 
production channel, especially in their evolution to higher 
$Q^2$ probes.

\mytextwidthfigure{t!}{
  \begin{tabular}{cc}
    \includegraphics{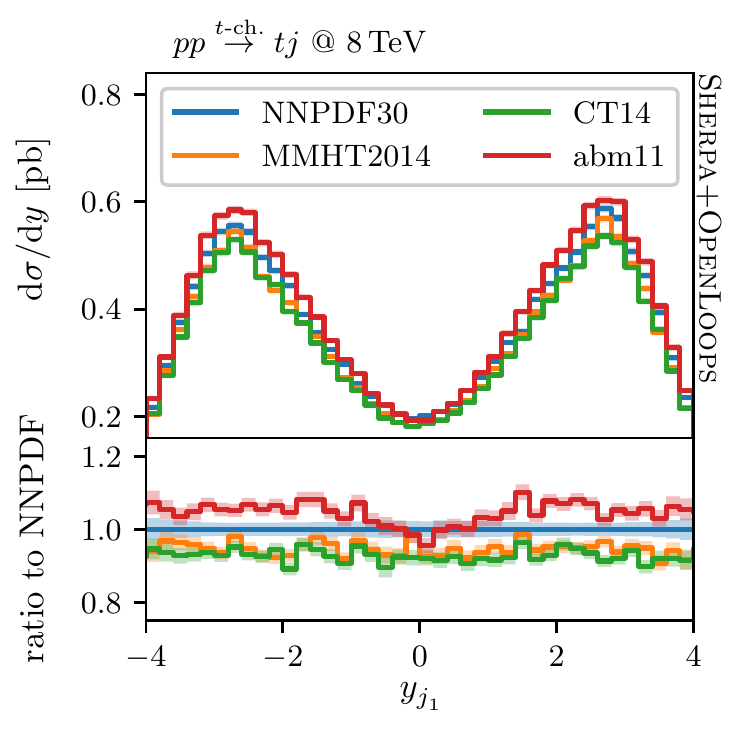} &
    \includegraphics{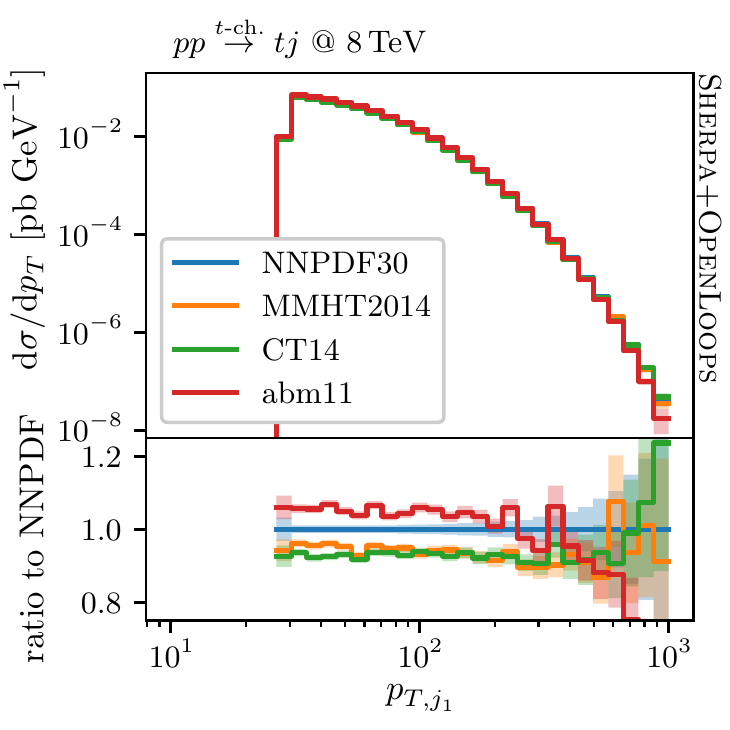} \\
    \includegraphics{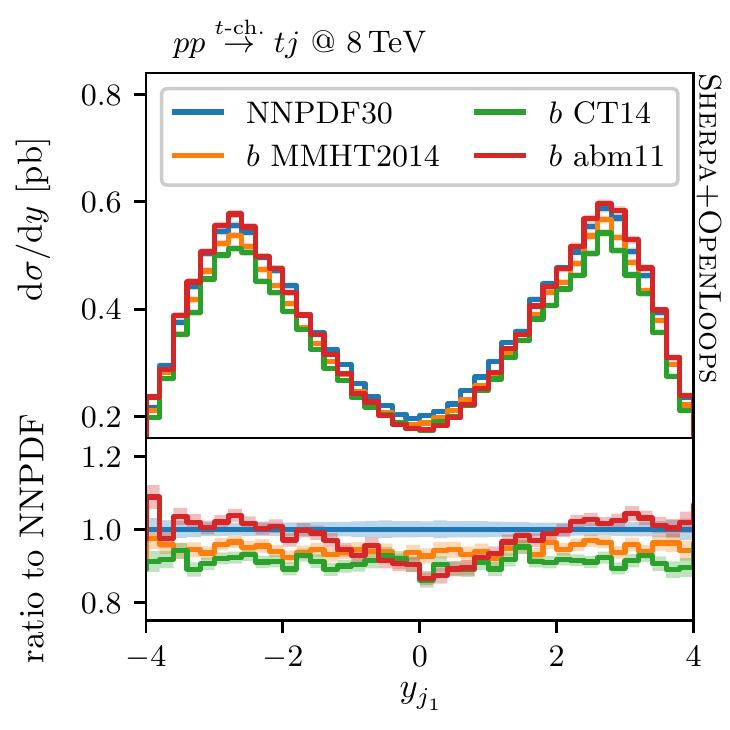} &
    \includegraphics{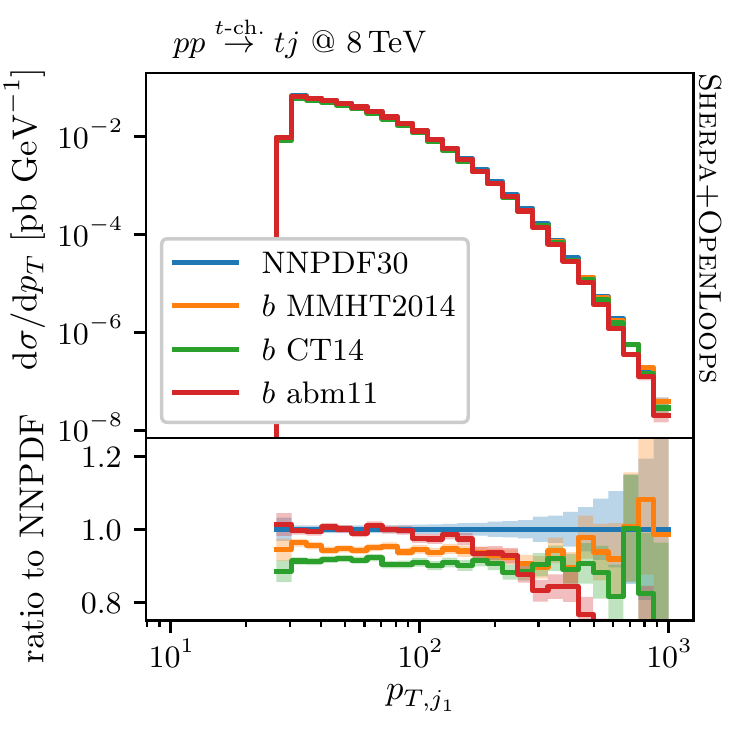}
  \end{tabular}
  }
  {\label{fig:PDF_JetDists}The impact of different PDF sets on the 
  leading-jet rapidity (left column) and transverse momentum (right column) in 
  \MCatNLO $t$-channel single-top production.  We show results for the 
  variation of all PDFs (top row) and of the bottom PDF only (bottom row). 
  The uncertainty band gives the statistical errors.}

\begin{figure}
  \begin{center}
    \includegraphics[width=0.60\textwidth]{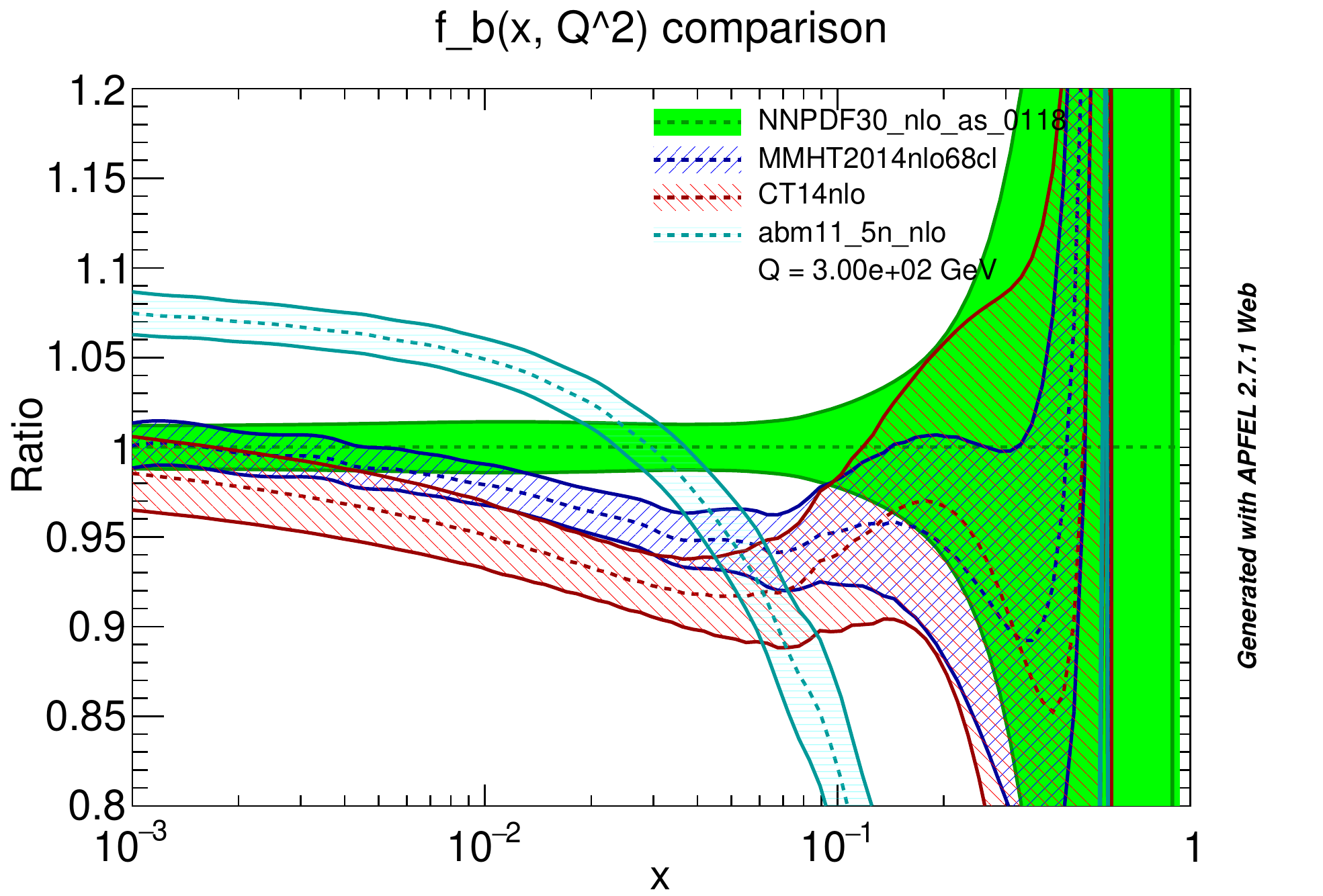}
    \caption{\label{PDF_Ratios}Ratios between different PDF sets for the bottom
    PDF $f_b(x, Q^2)$, with the scale $Q$ set to the average factorisation
    scale for $t$-channel single-top production.  The plot has been generated
    using the \APFEL library~\cite{Bertone:2013vaa}.}
  \end{center}
\end{figure}

This highlights that for a successful extraction of $|V_{tb}|$ from 
single-top production a good understanding of the bottom-quark PDF 
and its uncertainty is required.%
\footnote{See~\cite{Alekhin:2015cza} for a discussion of the \emph{light}-quark
PDF dependence of $t$-channel single top-quark production.}
Whereas errors internal to a PDF set are usually taken into account for such a
measurement~\cite{Aaboud:2017pdi}, the spread over different PDF sets should
also be included, as is done in~\cite{Khachatryan:2014iya}.  Instead of
extracting $|V_{tb}|$, single-top production cross sections can also be used
to fit the bottom-quark PDF, assuming $|V_{tb}|\approx 1$.  To explore this,
we study both total and differential cross sections for $t$-channel 
single-top production
varying the input PDF set, comparing central values of \NNPDF\,3.0,
CT14~\cite{Alekhin:2012ig}, MMHT2014~\cite{Harland-Lang:2014zoa} and
abm11~\cite{Alekhin:2012ig}, all at NLO.  To capture correlation effects
we vary either all parton densities, or the bottom-quark density only
while leaving the other densities at their default \NNPDF\,3.0 values. 

\FigureRef{fig:PDF_JetDists} shows the leading jet rapidity $j_{y_1}$ (left
column) and its transverse momentum $p_{T,j_1}$ (right column), in the fiducial
region of the ATLAS measurement \cite{Aaboud:2017pdi}.
In the top row, the PDF set is varied for all flavours, whereas in the
bottom row, only the bottom and anti-bottom PDF is varied.
In summary, 
when varying among the \NNPDF\,3.0, the CT14 and the MMHT2014 sets, we find a
mostly flat ratio between the rates and practically all relevant distributions,
with CT14 and MMHT2014 approximately \SI{5}{\percent} below \NNPDF\,3.0.  This
finding does not change much when only varying the bottom PDF, suggesting that
the normalisation is driven by the respective bottom-quark densities, with
the lighter quark and gluon densities agreeing among the PDF sets in the
relevant phase-space regions.
This is consistent with the ratios between the bottom PDFs in
\FigRef{PDF_Ratios}, where they are shown at the scale $Q=\SI{300}{\GeV}$. This
is approximately the average factorisation scale for our $t$-channel single-top
production.  The distribution of the longitudinal momentum fraction peaks at
$x\approx5\cdot10^{-3}$, with an average of $x\approx10^{-1}$. In this region
we indeed find the MMHT2014 and CT14 bottom PDF values to be
5--\SI{10}{\percent} smaller than the \NNPDF\,3.0 ones.
For abm11, we observe in \FigRef{fig:PDF_JetDists} for top production a
normalisation offset with respect to \NNPDF\,3.0 of about $+\SI{5}{\percent}$,
which however completely vanishes
when only varying the bottom-quark density.
In addition, we find shape dependences at the level
of \SI{10}{\percent} for the leading-jet rapidity distribution $y_{j_1}$. 
Similarly, the leading jet transverse momentum exhibits strongly divergent 
shapes of a similar magnitude as the five- to four-flavour calculation 
difference beyond $\pt>200\,\text{GeV}$. 

These findings suggest that in order to improve the bottom-quark distribution 
from single-top production, its measurement at higher luminosities and/or 
energies is mandatory such that both the high-$\pT$ and central rapidity 
regions can be explored with competitive statistical uncertainties. 
In turn, this implies, despite the observed differences, 
that the bottom-quark PDF uncertainty is sufficiently well understood 
for $|V_{tb}|$ extractions from single-top production with the \SI{8}{\TeV} data.

\section{Signal-over-background ratio for different light-jet cuts}
\label{Sec::Results}

Finally, we investigate the impact and effectiveness of different 
particle-level cuts to enhance the $t$-channel single-top signal
over the background, consisting of $W$-boson and 
$t\bar{t}$+jets production.
To this end, we are analysing our particle-level calculation using 
the Rivet analysis framework~\cite{Buckley:2010ar}.  
The particle reconstruction and the cuts we apply to our generated 
samples are chosen such as to emulate the analysis strategy and 
object definitions used for Monte-Carlo samples in a recent 
experimental single-top study~\cite{Aaboud:2017pdi},
except for the light-jet multiplicity cut as described below.

First, leptons are dressed with all photons within a radius $R=0.1$, 
and then are required to have $|\eta_\ell| < 2.5$ and 
$p_T^\ell > \SI{25}{GeV}$. 
Dressed leptons that do not originate from any hadron decay 
(either directly or via an intermedia $\tau$ lepton decay) are 
then considered to be tagged leptons.  We require exactly one tagged lepton.
In the setups we use this is guaranteed implicitly:
Any tagged lepton $\ell$ is generated via $W \rightarrow \ell$ or
$W\rightarrow \tau \rightarrow \ell$. 
We further require a missing transverse momentum, $p_T^\text{miss}$, 
of at least $\SI{30}{GeV}$. 
The $W$ boson is then reconstructed from the tagged (dressed) lepton momentum
and $p_T^\text{miss}$, using $m_W$ as a constraint.
Jets are defined by the \antikT-algorithm~\cite{Cacciari:2008gp} with a radius
parameter $R = 0.4$, $p_T > \SI{30}{GeV}$ and $|\eta|<4.5$. 
They are built from all visible particles except for dressed leptons. 
If one of the jets lies within $R = 0.4$ around the tagged lepton, 
the event is vetoed.  
Jets are tagged as $b$-jets by associating a $b$-hadron with a 
ghost-matching method~\cite{Cacciari:2008gn}, and if their 
pseudo-rapidity is $|\eta| < 2.5$.  
Exactly one $b$-jet is required. 
Events are rejected if $m_{b\ell} > \SI{160}{GeV}$ to stay 
away from the off-shell regions.  
The top quark is then reconstructed by adding the four-momenta of the 
reconstructed $W$ boson and the $b$-jet.  
The remaining jets are called light jets (or ``l-jets'') and we number them
according to their transverse momenta as $j_n$, with $j_1$ being the leading
jet.

To further reduce the background, \cite{Aaboud:2017pdi} additionally requires
that there is exactly one light jet ($j_1$ in our notation),
$N_\text{l-jets}=1$. 
We study various alternatives for this cut in the following and assess 
their effectiveness. 

Focusing on the dominant $t$-channel production mode it is worthwhile
to contemplate its kinematics.  It is defined by the exchange of a
colourless $W$ boson in the $t$-channel, giving rise to a light ``tag'' jet
and the top quark which decays into a bottom-quark and a $W$ boson, with the latter
subsequently decaying either into a lepton-neutrino or a quark-\-anti-quark
pair.  The colourless $t$-channel exchange suggests a kinematic similarity
with weak-boson fusion events, with one light-quark current connected
to a heavy-quark current---the transition from bottom- to top-quark at Born
level. 
This analogy implies that, while the top quark and its decay system 
remain nearly inert in the central region of the detector, the 
light tag jet is peaked in the forward regions, at rapidities 
of about or above $|y_j|\approx 2$, cf.\ \FigRef{fig:s-vs-t}.
The contribution of the up-quark to the total 
top-quark production is larger than that of the down-quark 
to the total antitop-quark production. Through its valence bump at 
comparably large momentum fractions of $x\approx 0.15$ the 
mean rapidity of the tag jet will be somewhat larger
for top than for anti-top production.
Indeed, we find $\langle|y_t|\rangle = 2.23$ and $\langle|y_{\bar{t}}|\rangle
= 2.03$ for top-quark and antitop-quark production, respectively, when using
$N_\text{l-jets}=1$ as the light-jet multiplicity cut.  This gives a difference
of $\Delta \langle|y|\rangle = 0.20$, a value that varies between
$0.15\ldots0.22$ with the other cuts given below.

Due to the coherence property of QCD, additional radiation off the light
quark line will typically also be quite forward, while radiation off the
top quark is massively reduced due to the shielding of the collinear
singularity by its mass.  Therefore, additional QCD radiation in
the central region will be depleted and mainly driven by secondary emissions
from the top decay.  This feature, depletion of radiation in the central
region and a ``rapidity gap'' between the reconstructed top and the tag
jet are absent in the backgrounds, which are not driven by colourless
$t$-channel exchanges, but are more or less exclusively driven by the strong
interaction between the two protons.  This opens up possibilities for
substantial improvements of the signal-to-background ratio through 
cuts on additional central hadronic or jet activity.
In the following we test the effect of applying five different vetoes on central
QCD radiation:
\begin{enumerate}[label=\alph*)]
\item a simple cut on the rapidity difference between the reconstructed top and
  any light-jet $j$,
  \begin{equation}
    \Delta y_{tj}\,=\,|y_{t}-y_{j}|\,>\,y_\text{cut}\,,
  \end{equation}
\item a cut on light-jet activity in the central region, by demanding
  \begin{equation}
    G_T^{(0)}\,=\,\sum\limits_{j\in{\text{jets}}}\,|p_{\perp,j}|\,
    \exp\left(-|y_j|\right)\,<\,G^{(0)}_{T,\text{cut}}\,,
  \end{equation}
\item a cut on light-jet activity in the region around the top, by demanding
  \begin{equation}
    G_T^{(t)}\,=\,\sum\limits_{j\in{\text{jets}}}\,|p_{\perp,j}|\,
    \exp\left(-|y_j-y_t|\right)\,>\,G^{(t)}_{T,\text{cut}}\,,
  \end{equation}
\item a cut on the hadronic activity in the central region, by demanding 
  \begin{equation}
    g_T^{(0)}\,=\,\sum\limits_{i\in{\text{tracks}}}\,|p_{\perp,i}|\,
    \exp\left(-|y_i|\right)\,<\,g^{(0)}_{T,\text{cut}}\,,
  \end{equation}
\item a cut on the hadronic activity in the region around the top, by demanding 
  \begin{equation}
    g_T^{(t)}\,=\,\sum\limits_{i\in{\text{tracks}}}\,|p_{\perp,i}|\,
    \exp\left(-|y_i-y_t|\right)\,>\,g^{(t)}_{T,\text{cut}}\,.
  \end{equation}  
\end{enumerate}
Here, we use the properties of charged tracks to characterise the 
hadronic activity. 
They are defined to have $|\eta| < 2.5$ and $p_T > \SI{400}{MeV}$ 
and we discard tracks that are within $R=0.4$ around the $b$-jet 
or within $R=0.1$ around the lepton.
In consequence, no jet or track that has been used to reconstruct 
the top-quark enters the sum in the definition of the measures b)--e).
While all five options enhance the contribution from topologies 
that exhibit rapidity gaps, they vary in their restrictiveness. 
Only option a) rejects \emph{all} configurations for which the leading jet and
the top-quark are too close in rapidity. 
In contrast, the other four options weigh the occurring 
radiation by their distance either from the centre of 
the detector, $G_T^{(0)}$ and $g_T^{(0)}$, or the reconstructed 
top-quark, $G_T^{(t)}$ and $g_T^{(t)}$. 
Options b) and d) therefore do not necessarily lead to a 
rapidity gap between the top-quark and the light jet, but instead to a 
gradual depletion of the hadronic activity in the central 
detector. 

\mytextwidthfigure{t!}{%
  \includegraphics{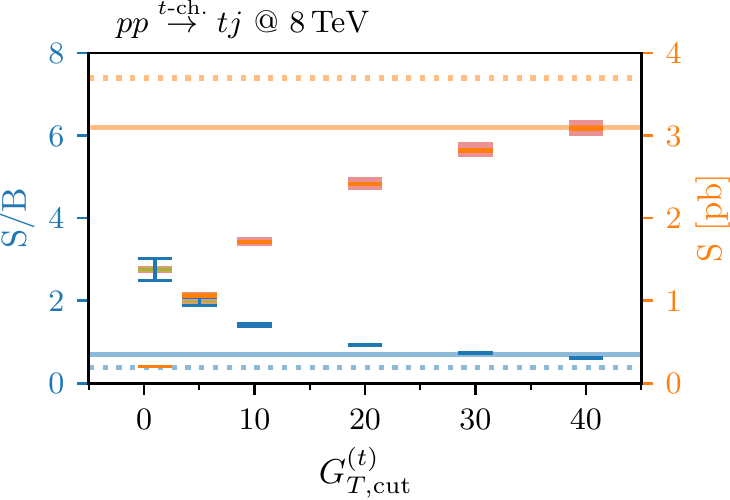}%
  \hfill
  \includegraphics{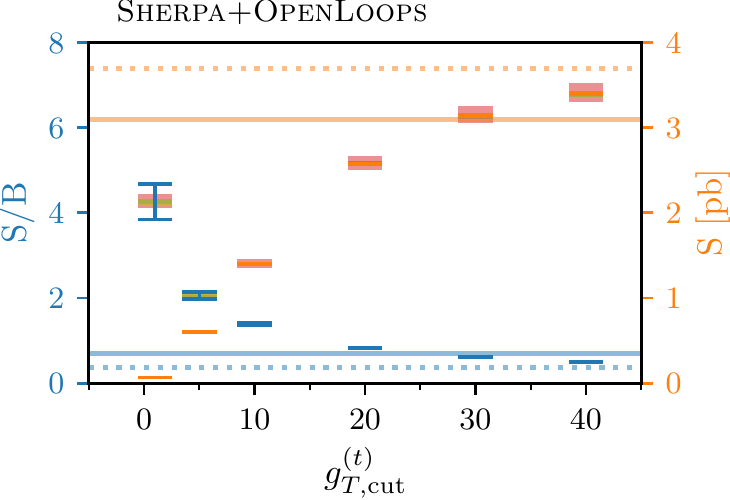}\\[2mm]
  \includegraphics{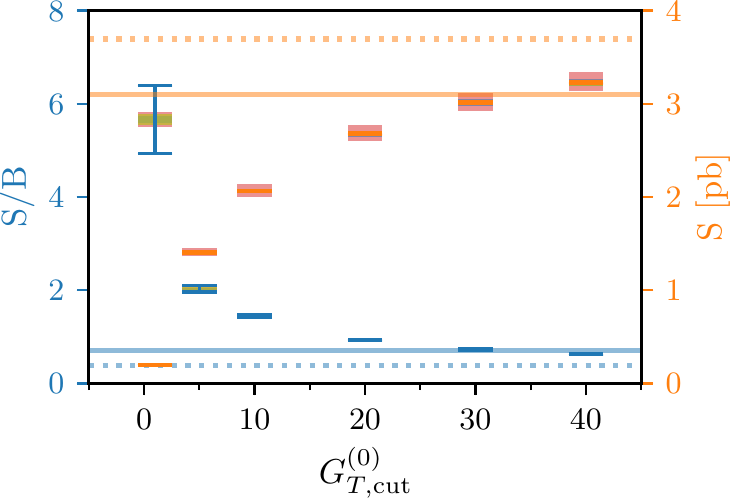}%
  \hfill
  \includegraphics{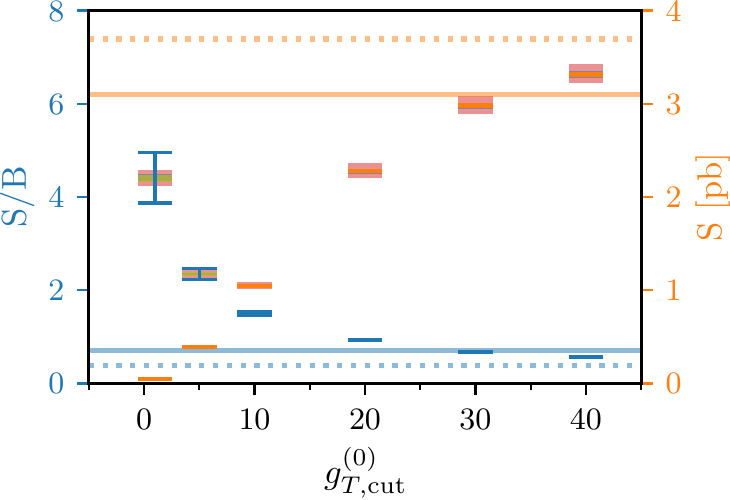}\\[2mm]
  \includegraphics{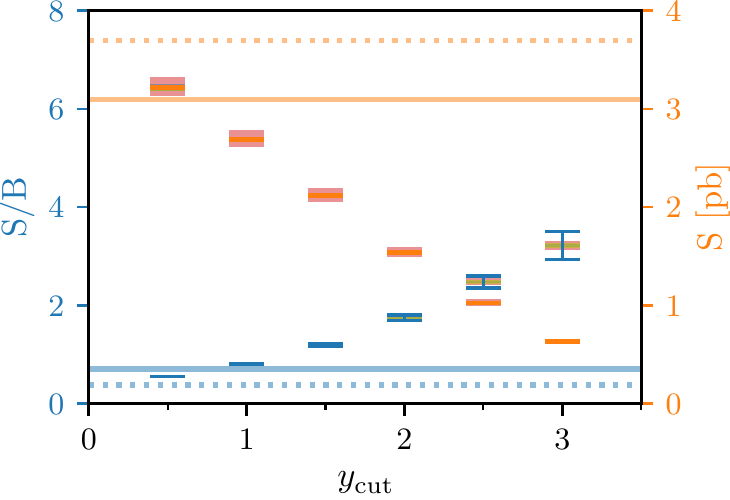}%
  \hfill
  \includegraphics{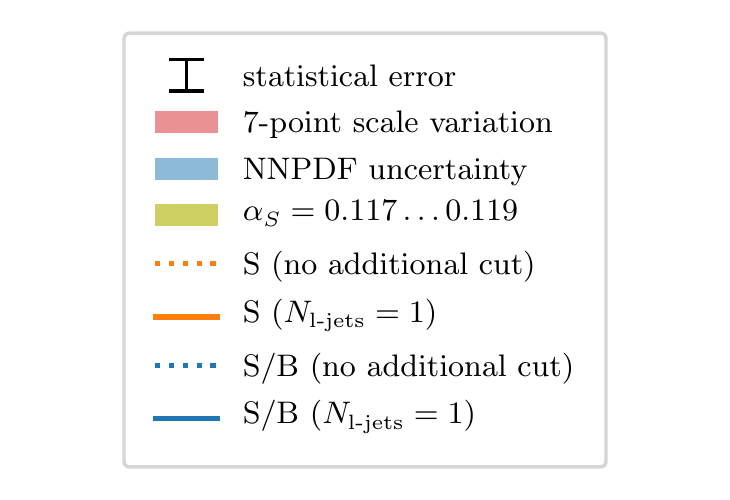}%
  }{\label{fig:sb}%
  Signal-over-background ratios for different veto cuts.  The signal is
  $t$-channel single-top production, the background consists of the sum over
  $t\bar{t}$ and $Wjb$ production. Scale uncertainties 
  are included only for the signal in the \SB\ ratios.
  PDF and \alphaS uncertainties are varied consistently for all calculations.
  The large statistical uncertainties visible for the highest \SB\ values
  originate in the considerable background suppression using the respective
  vetoes. We contrast the \SB\ ratios with the one resulting from the
  $N_\text{l-jets}=1$ requirement used in the original experimental analysis.}

\mytextwidthfigure{p}{%
  \includegraphics{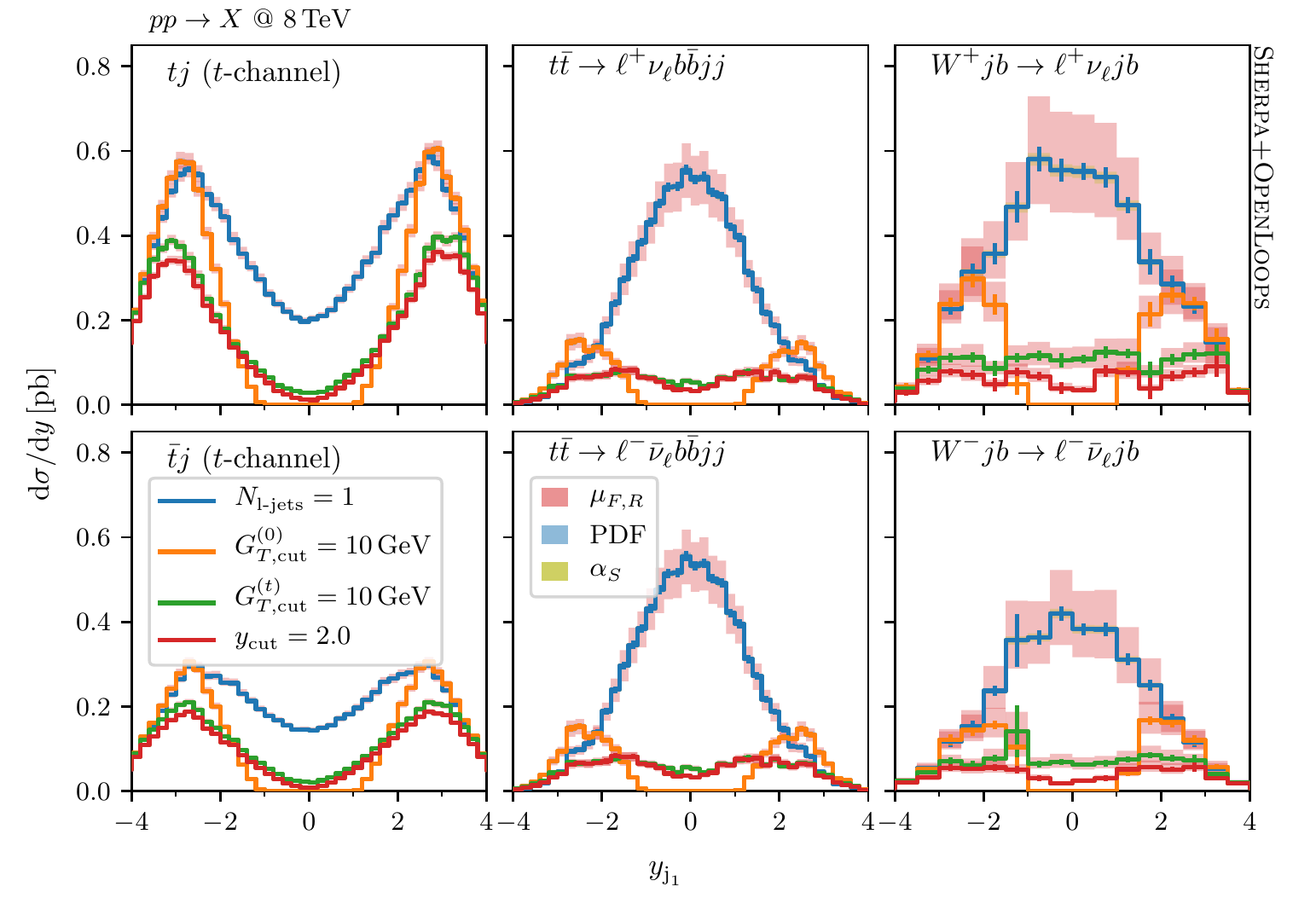}\\
  \includegraphics{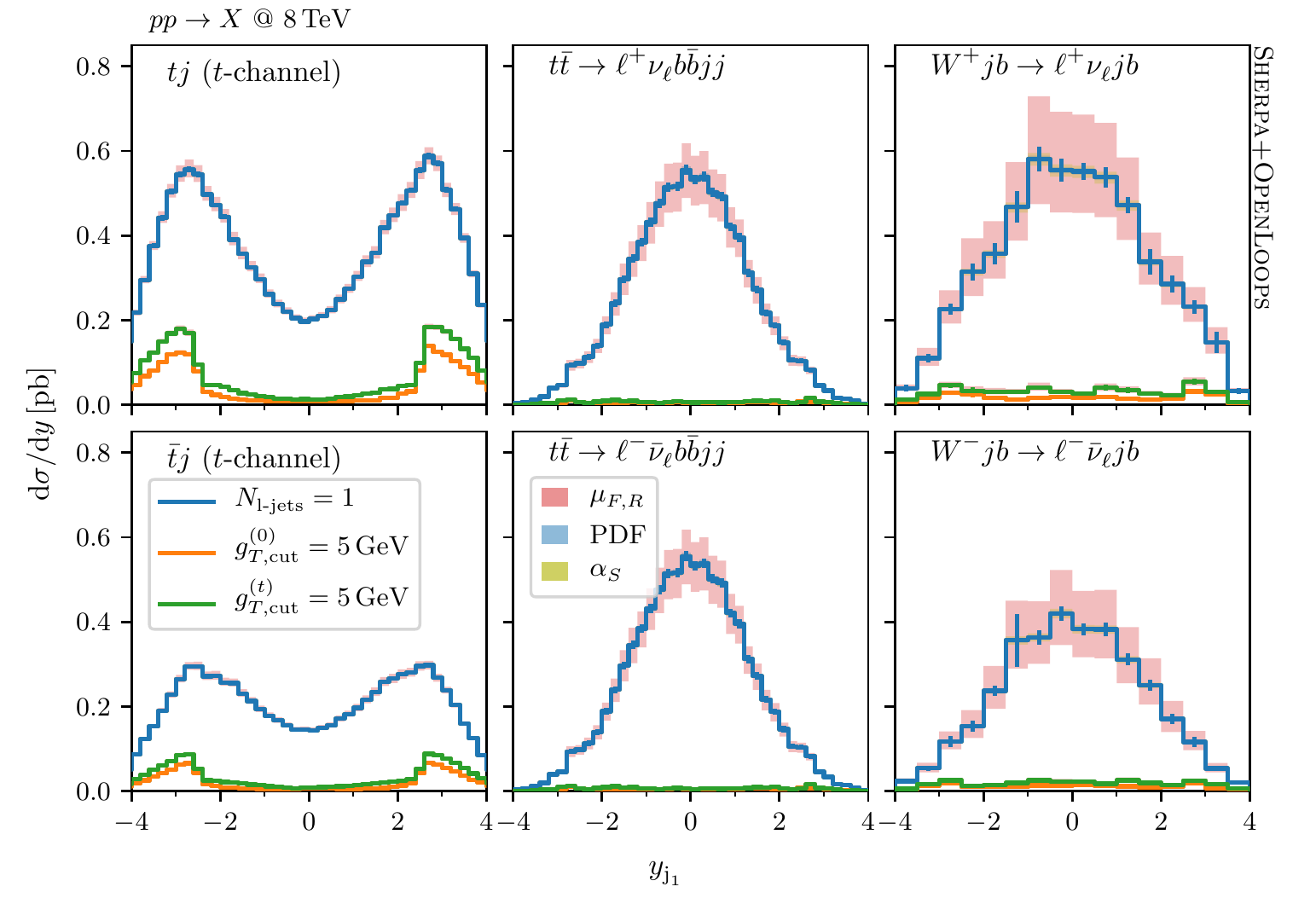}
  }{\label{fig:y}%
  \Sherpa \MCatNLO results for the leading-jet rapidity $y_{j_1}$ for different signal
  channels and background processes, given different veto cuts.}

The signal-over-background (\SB) ratios for all five versions 
of inducing a rapidity gap and the remaining signal cross 
sections are shown in \FigRef{fig:sb}. 
While \SB\ ratios of 4 and higher can be achieved, they of 
course come at the cost of a vanishingly small signal cross section. 
The best results can be obtained by restricting central-jet 
or hadronic activity. 
While the \SB\ ratios are similar between the two approaches, 
the jet-based veto removes less signal cross section than 
the track-based one and is therefore preferable. 
Interestingly, demanding a depleted central detector 
achieves at least as good and in most cases better results, 
than a depletion in a rapidity region relative to the 
reconstructed top quark.  This is true both in terms of \SB\ ratios and of the
remaining signal cross section.
Of course, since the top-quark itself is predominantly produced 
very centrally, the differences are moderate.
The track-based rejections fare very similarly to the jet-based 
rejections for large rejection scales, but are more repressive 
at small rejection scales. 
While backgrounds are suppressed very well, also the signal 
cross section is lost.
Good compromises are offered by central-jet or track veto 
scales of around 5--\SI{10}{GeV} or rapidity gaps of $2.5$ 
units.

To examine the effect of the above rapidity gap inducing 
phase space restrictions on the leading light-jet rapidity,
we define four sets of cuts:\\
\begin{minipage}{0.47\textwidth}
\begin{enumerate}[label=\roman*)]
  \item $G_{T,\text{cut}}^{(0)}=\SI{10}{GeV}\,,$
  \item $G_{T,\text{cut}}^{(t)}=\SI{10}{GeV}\,,$ 
\end{enumerate}
\end{minipage}
\hfill
\begin{minipage}{0.47\textwidth}
\begin{enumerate}[label=\roman*)]
  \setcounter{enumi}{2}
  \item $G_{T,\text{cut}}^{(0)}=\SI{5}{GeV}\,,$ 
  \item $G_{T,\text{cut}}^{(t)}=\SI{5}{GeV}\,.$ 
\end{enumerate}
\end{minipage}\\[1.0em]
The resulting distributions are shown in \FigRef{fig:y}. 
The upper panel shows the leading light-jet 
rapidity distributions for $t$-channel top and anti-top 
production after the application of i) or ii) and 
contrast them with the distributions after the application of
the original $N_\text{l-jets}=1$ or $y_\text{cut}=2.5$ 
restrictions.
They are accompanied by the corresponding distributions 
of the $t\bar{t}$ and $Wjb$ background 
processes. 
The lower panel shows the same distributions, now applying 
the restrictions of iii) and iv) instead.

Because the $N_\text{l-jets}=1$ requirement does not enforce a rapidity gap, we
find that when using this requirement a large number of background events
survive where the signal cross section is minimal.
Among the three other options, $G_{T,\text{cut}}^{(t)}=\SI{10}{GeV}$ 
and $y_\text{cut}=2.5$ largely give very similar results, 
with minor difference in the central region for the $Wjb$ 
background. 
$G_{T,\text{cut}}^{(t)}=\SI{10}{GeV}$ induces a more 
aggressively depleted central detector, but leaves 
an increased signal rate at $|y_{j_1}|\approx 2.5$, 
accumulating to a larger signal cross section throughout 
the spectrum.
Decreasing the track-veto scale in the central region or 
the vicinity of the top-quark reduces the background 
rates to negligible values, but also has an adverse 
effect on the signal cross section, as we have already observed before.
Only small signal regions beyond $|y_{j_1}|\gtrsim 2.5$ 
survive.

Turning the above line-of-thought around, requiring a minimal 
remaining signal cross-section after cuts of \SI{1}{pb}, 
the best value for \SB\ using a plain rapidity gap requirement 
is about 2.5 when using $y_\text{cut}=2.5$.  
While the light-jet suppression in the top vicinity 
achieves similar results, leaving a signal cross section of 
\SI{1}{pb} with a \SB\ of about 2.4 with 
$G_{T,\text{cut}}^{(t)}=\SI{3.5}{GeV}$,
the top-independent central jet veto performs best, reaching 
a \SB\ ratio of approximately 4 at a signal cross section 
of \SI{1}{pb} with $G_{T,\text{cut}}^{(0)}=\SI{3}{GeV}$.

\section{Summary}
\label{sec:summary}
We reported on the simulation of single top-quark production in
the $s$-, $t$- and $tW$-channels with the \Sherpa event generator 
at \MCatNLO accuracy.  After
validating our results with experimental data for various cross sections
and through selected differential observables, we focused on two short
phenomenological studies.  First, we analysed the impact of the bottom PDF
on various observables.  We find that for most of the standard PDFs the
shapes are very robust, on the level of 5\% or below, and that the main
differences are in the total normalisation, i.e.\ the overall cross section
with bottom PDF induced uncertainties of up to about 10\%.  
The only exception is the abm11 PDF set, which also
shows some shape distortions.  Overall, this provides ample motivation to use
precision determinations of single top-quark production as a means to directly
measure the absolute value of the CKM element $|V_{tb}|$.  Second, we focused on
the $t$-channel production mode and applied a variety of vetoes on
QCD radiation in central rapidity regions.  To this end we introduced a
number of observables, essentially scalar sums of transverse momenta of jets
or charged tracks, weighted with an exponential form suppressing them at
large rapidities or rapidity differences with respect to the top-quark system.  
As they exploit the topological differences of the signal and its 
background processes, it is unsurprising that all five versions of such 
an additional requirement provided 
significant enhancements of the $S/B$ ratio of around 2-4 while simultaneously 
keeping the signal cross section at \SI{1}{pb} or above.
This leads us to suggest to replace the flat restriction on 
light-jet activity used so far in experimental analysis by any of the 
rapidity gap inducing candidates suggested in this paper and investigate 
their behaviour further in subsequent experimental studies.


\section*{Acknowledgements}
This work was supported by the European Union as part of the EU Marie
Curie Research Training Network MCnet (MRTN-CT-2006-035606).
We thank all members of the \Sherpa collaboration for valuable input 
and S. H\"oche in particular for providing help with the \Comix matrix 
element generator in algorithmically differentiating the three 
different single-top processes in NLO calculations.

\appendix

\bibliographystyle{bib/amsunsrt_modp}
\bibliography{bib/journal}

\end{document}